\documentclass[final,1p,times]{elsarticle}

\usepackage{graphicx}
\usepackage{amssymb}
\usepackage{amsthm}
\usepackage{lineno}
\usepackage{amsmath}
\usepackage{amssymb}

\newcommand{\sqrtsNN}{\mbox{$\sqrt{\mathrm{\it s_{NN}}}$} }
\newcommand{\vtwo}{$v_{2}$ }

\newcommand{\ks}{${K}^{0}_{S}$ }

\newcommand{\lam}{$\Lambda$ }

\def \auau  {Au+Au }

\def \GeVc {\mbox{$\mathrm{GeV}/c$}}

\def \lt {\mbox{$<~$} }

\journal{Nuclear Physics A}

\begin{document}

\begin{frontmatter}

\title{Event anisotropy $v_{2}$ in Au+Au collisions at $\sqrt{s_{NN}}=$ 7.7 - 62.4 GeV with STAR}


\author{Shusu Shi$^{a, b}$ (for the STAR Collaboration)}

\address[a]{Institute of Particle Physics, Huazhong
Normal University, Wuhan, Hubei, 430079, China}
\address[b]{Key Laboratory of Quark and Lepton Physics (MOE) and Institute of Particle
Physics, Central China Normal University, Wuhan 430079, China
}

\begin{abstract}
We present the $v_2$ measurement at midrapidity from Au+Au collisions at $\sqrt{s_{NN}}=$ 7.7, 11.5, 19.6, 27, 39 and 62.4 GeV for inclusive charged hadrons and identified hadrons
($\pi^{\pm}$, $K^{\pm}$, $K_{S}^{0}$, $p$, $\bar{p}$, $\phi$, $\Lambda$, $\bar{\Lambda}$,
$\Xi^{-}$, $\bar{\Xi}^{+}$, $\Omega^{-}$, $\bar{\Omega}^{+}$) up to 4 GeV/$c$ in $p_{T}$.
The beam energy and centrality dependence of charged hadron $v_2$ are presented with comparison to higher energies at RHIC and LHC.
The identified hadron $v_{2}$ are used to discuss the NCQ scaling
for different beam energies. Significant difference in $v_{2}(p_{T})$ is observed between
particles and corresponding anti-particles for $\sqrt{s_{NN}} <$ 39 GeV.
These differences are more pronounced for baryons
compared to mesons and they increase with decreasing energy.

\end{abstract}

\end{frontmatter} 


\section{Introduction}
Searching for the phase boundary in the Quantum ChromoDynamics (QCD) phase diagram is one of the main motivations
of the Beam Energy Scan (BES) program at RHIC. The elliptic flow ($v_2$) could be used as a powerful tool~\cite{review},
because of the sensitivity of underlying dynamics in the early stage of the
collisions. The Number of Constituent Quark (NCQ) scaling in the top energy collisions at RHIC (\sqrtsNN = 200 GeV)
indicates the collectivity has been built up in the partonic level ~\cite{starklv2, flowcucu}.
Recently, the similar NCQ scaling of multi-strange hadrons,
$\phi$ and $\Omega$, which are less sensitive to the late hadronic interactions provides the clear evidence of partonic collectivity~\cite{phiomega}.
An energy dependence study based on A Multi-Phase Transport (AMPT) model indicates the NCQ scaling is related to the degrees of freedom in the system~\cite{AMPTNCQ}.
If the partonic degree of freedom is included in the AMPT model, the NCQ scaling (including multi-strange hadrons) could be observed;
whereas the NCQ scaling is broken in the case of including only hadronic degree of freedom.
The BES data offer us the opportunity to investigate
the QCD phase boundary with $v_2$ measurements.
In this paper, we present the \vtwo results
from the STAR experiment in Au+Au collisions at \sqrtsNN = 7.7 - 62.4 GeV.
The particle identification for $\pi^{\pm}$, $K^{\pm}$ and $p~(\overline{p})$ is
achieved via the energy loss in the Time Projection Chamber (TPC)~\cite{STARtpc} and the time of flight information from the multi-gap resistive plate chamber detector~\cite{STARtof}.
Strange and multi-strange hadrons are reconstructed with the decay channels:
\ks $\rightarrow \pi^{+} + \pi^{-}$, $\phi \rightarrow K^{+} +
K^{-}$, \lam $\rightarrow p + \pi^{-}$ ($\overline{\Lambda}
\rightarrow \overline{p} + \pi^{+}$),
$\Xi^{-} \rightarrow$ \lam $+\ \pi^{-}$ ($\overline{\Xi}^{+}
\rightarrow$ $\overline{\Lambda}$+\ $\pi^{+}$)
and $\Omega^{-} \rightarrow$ \lam $+\ K^{-}$ ($\overline{\Omega}^{+}
\rightarrow$ $\overline{\Lambda}$+\ $K^{+}$).
The event plane method~\cite{v2Methods1} and cumulant method~\cite{cumulant1, cumulant2} are used for the $v_{2}$ measurement.

\section{Results and Discussions}

\begin{figure*}[ht]
\vskip 0cm
\begin{center} \includegraphics[width=0.7\textwidth]{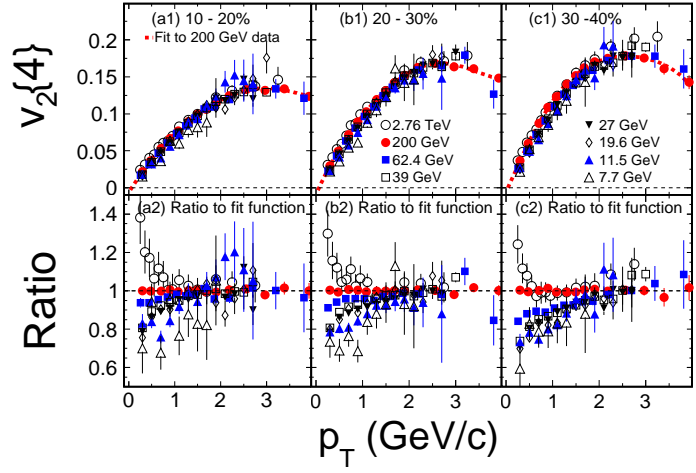}\end{center}
\caption{(Color online) The top panels show $v_2\{4\}$ vs. $p_T$ at midrapidity for various collision energies (\sqrtsNN = 7.7 GeV to 2.76 TeV).
The results for \sqrtsNN = 7.7 to 200 GeV are for \auau collisions and those for 2.76 TeV are for Pb + Pb collisions.
The dashed red curves show the fifth order polynomial fits to the results from \auau collisions at \sqrtsNN = 200 GeV. The bottom panels show
the ratio of $v_2\{4\}$ vs. $p_T$ for all \sqrtsNN with respect to the fit curve. The results are shown for
three collision centrality classes: $10 - 20\%$ (a1), $20 - 30\%$ (b1) and $30 - 40\%$ (c1)~\cite{BESchv2}. Error bars are shown only for the statistical uncertainties respectively.}
\label{v2_4_pt_beam_energy}
\end{figure*}

\begin{figure*}[ht]
\vskip 0cm
\begin{center}\includegraphics[width=0.7\textwidth]{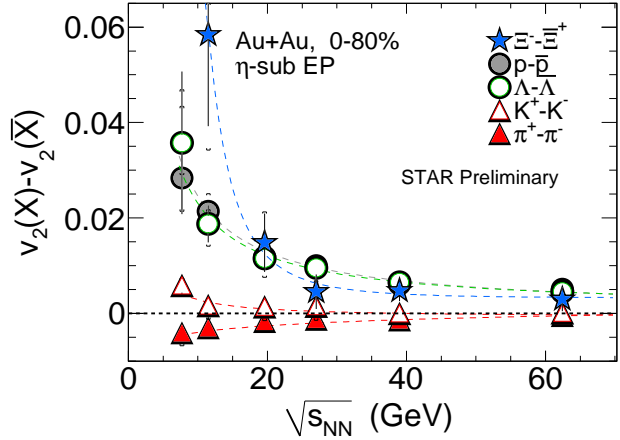}\end{center}
\caption{(Color online) The difference in $v_2$ between particles and their corrsponding anti-particles ($v_{2}(X)-v_{2}(\bar{X})$)
as a function of beam energy in Au+Au collisions (0-80\%). The statistical and systematic uncertainties are shown by vertical line and cap respectively.
The dashed lines in the plot are fits with the equation described in the text.}
\label{v2_diff}
\end{figure*}

\begin{figure*}[ht]
\vskip 0cm
\includegraphics[width=1.0\textwidth]{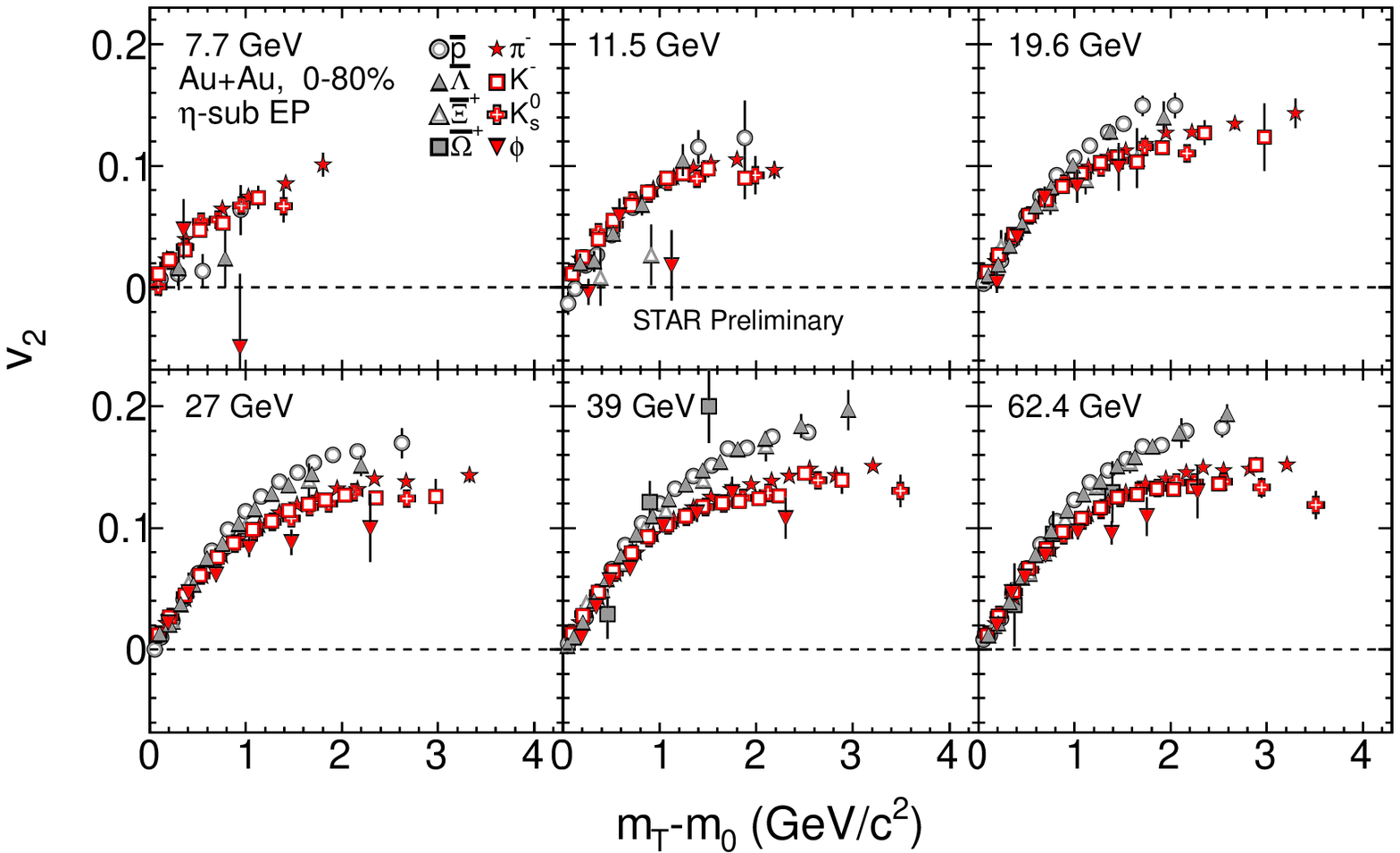}
\caption{(Color online) The elliptic flow ($v_2$) as a function of transverse mass ($m_{T} - m_{0}$) for the selected particles in Au+Au collisions (0-80\%) at \sqrtsNN = 7.7, 11.5, 19.6, 27, 39 and 62.4 GeV. Error bars are shown only for the statistical uncertainties.}
\label{mt_m}
\end{figure*}

Figure~\ref{v2_4_pt_beam_energy}~\cite{BESchv2} shows the transverse momentrum ($p_T$) dependence of $v_2\{4\}$ from
\sqrtsNN = 7.7 GeV to 2.76 TeV in $10 - 20\%$ (a1), $20 - 30\%$ (b1) and $30 - 40\%$ (c1) centrality bins. The
ALICE results in Pb + Pb collisions at $\sqrt{s_{NN}}$ = 2.76 TeV are taken from Ref.~\cite{alicev2}.
The 200 GeV data is empirically fit by a fifth order polynomial function.
For comparison, the $v_2$ from other energies are divided by the fit and shown in the lower panels of
Fig.~\ref{v2_4_pt_beam_energy}.
For $p_T$ below 2~\GeVc, the $v_2$ values rise with increasing collision energy. Beyond $p_T = 2~\GeVc$
the $v_2$ results show comparable values within statistical errors.
The increase of $v_{2}(p_{T})$ as a function of energy could be
due to the change of chemical composition from low to high energies and/or larger collectivity at higher collision energy.

Figure~\ref{v2_diff} shows the $p_T$ independent difference in $v_2$ between particles and their corresponding anti-particles. The data of collisions at \sqrtsNN= 19.6 and 27 GeV is new since Quark Matter 2011~\cite{qm2011}. The $\eta$-sub event plane method is used for the measurement. In this method, one
defines the flow vector for each particle based on particles
measured in the opposite hemisphere in pseudorapidity ($\eta$).
An $\eta$ gap of $|\eta| < 0.05$ is used between negative/positive $\eta$ sub-event
to reduce the non-flow effects by enlarging the separation
between the correlated particles.
$v_{2}(X)-v_{2}(\bar{X})$ ($\Delta v_{2}$) represents
the average values of the difference in $v_2$ between particles and corresponding anti-particles over measured $p_{T}$ range.
The dashed lines in the Fig.~\ref{v2_diff} are fits with the function:
$f_{\Delta v_{2}} (\sqrt{s_{NN}}) = a\sqrt{s_{NN}}^{-b}+c$. A monotonic increase of $\Delta v_{2}$ with decreasing collision energy is observed
and the slope of the difference increases towards lowers energies.
The difference is more pronounced for baryons
compared to mesons.
The observed difference in $v_2$ reflects that the particles vs. anti-particles could
not fit into the NCQ scaling. The breaking of NCQ scaling between particles and anti-particles indicates the contributions from hadronic interactions increase in the system evolution with decreasing collision
energy.
The energy dependence of $\Delta v_{2}$ could be qualitatively reproduced by the baryon transport effect~\cite{transporteffect} or hadronic potential effect~\cite{hadronicpotential}. So far theoretical calculations fail to
quantitatively reproduce the measured $v_2$ difference and none of the calculations
could explain the correct order of particles.

Figure~\ref{mt_m} shows the $v_2$ as a function of transverse mass ($m_{T}-m_{0}$) for the selected particles for all six collision energies. In the
top energy (\sqrtsNN= 200 GeV) collisions, a clear splitting in $v_2$ between baryons and mesons is observed for $m_{T}-m_{0} > 1~ {\rm GeV}/c^{2}$. The splitting between baryons and mesons suggest the
system created in the collisions is sensitive to the quark degree of freedom. The selected particles show a similar splitting for collision energy $\geq$  39 GeV. The bayron and meson groups become closer to each other at all lower energies. At \sqrtsNN= 11.5 GeV,
the splitting between baryons and mesons is almost gone. The clear trend, a decreasing baryon-meson splitting of $v_{2}(m_{T}-m_{0} )$ beyond $m_{T}-m_{0} > 1~ {\rm GeV}/c^{2}$ indicates the hadronic interactions
become more important in the lower collision energies.

\section{Summary}
In summary, we present the $v_2$ measurements for charged hadrons and identified hadrons in Au+Au collisions at
\sqrtsNN= 7.7 - 62.4 GeV. The comparison with \auau collisions at higher energies at RHIC (\sqrtsNN = 62.4 and 200 GeV)
and at LHC (Pb + Pb collisions at \sqrtsNN = 2.76 TeV) shows the $v_2\{4\}$ values at low $p_T$ ($p_T <$ 2.0 GeV/$c$) increase with increase in
collision energy.
The baryon and anti-baryon $v_2$ show significant difference for \sqrtsNN $\lt$ 39 GeV.
The difference of $v_2$ between
particles and corresponding anti-particles (pions, kaons, protons, $\Lambda$s and $\Xi$s)
increases with decreasing the beam energy.
The baryon-meson splitting of $v_{2}(m_{T}-m_{0} )$ beyond $m_{T}-m_{0} > 1~ {\rm GeV}/c^{2}$ becomes
smaller in the lower collisions energy and is almost gone in collisions at \sqrtsNN = 11.5 GeV.
Experimental data indicate that the hadronic interactions become more important at the lower
beam energy.

\section{Acknowledgments}
This work was supported in part by the National Natural Science Foundation of China under grant No. 11105060, 10775060, 11135011, 11221504 and self determined
research funds of CCNU from the colleges' basic research and operation of MOE.

\end{document}